\documentclass[twocolumn,conference]{IEEEtran}
\usepackage[T1]{fontenc}
\usepackage[latin9]{inputenc}
\usepackage{algorithm2e}
\usepackage{amsthm}
\usepackage{amsmath}
\usepackage{amssymb}
\usepackage{graphicx}
\usepackage{esint}

\makeatletter
  \theoremstyle{plain}
  \newtheorem{thm}{\protect\theoremname}
  \theoremstyle{plain}
  \newtheorem{lem}{\protect\lemmaname}
  \theoremstyle{plain}
  \newtheorem{prop}{\protect\propositionname}

\providecommand{\lemmaname}{Lemma}
\providecommand{\propositionname}{Proposition}

\providecommand{\theoremname}{Theorem}

\usepackage[draft]{hyperref}

\makeatother

\providecommand{\lemmaname}{Lemma}
\providecommand{\propositionname}{Proposition}
\providecommand{\theoremname}{Theorem}

\begin{document}

\title{Compute-and-Forward over Block-Fading Channels\\
 Using Algebraic Lattices}

\author{\IEEEauthorblockN{Shanxiang Lyu, Antonio Campello and Cong Ling}
\IEEEauthorblockA{Department of EEE, Imperial College London\\
 London, SW7 2AZ, United Kingdom \\
 Email: s.lyu14, a.campello, c.ling@imperial.ac.uk}\and \IEEEauthorblockN{Jean-Claude Belfiore} \IEEEauthorblockA{Mathematical and Algorithmic Sciences Lab\\
 France Research Center\\
 Huawei Technologies\\
 belfiore@telecom-paristech.fr}}
\maketitle
\begin{abstract}
Previous approaches to compute-and-forward (C\&F) are mostly based
on quantizing channel coefficients to integers. In this work, we investigate the C\&F strategy
over block fading channels using Construction A over rings, so as to allow better quantization for
the channels. Advantages in decoding error probabilities and computation
rates are demonstrated, and the construction is shown to outperform the C\&F strategy over the integers
$\mathbb{Z}$. \end{abstract}


\section{Introduction}

Building upon the property that lattice codes are closed under integer
combinations of codewords, the compute-and-forward (C\&F) relaying
protocol proposed by Nazer and Gaspar \cite{Nazer2011} has become
a popular physical layer network coding framework. The protocol has
been extended in several directions. Since $\mathbb{Z}$ may not be
the most suitable space to quantize the actual channel, one line of
work is to use more compact rings. If the message space and the lattice
cosets are both $\mathcal{O}$-modules where $\mathcal{O}$ refers
to a ring, the linear labeling technique in \cite{Feng2013a} enables
the decoding of a ring combination of lattice codewords. It has also
been shown that using Eisenstein integers $\mathbb{Z}[\omega]$ \cite{Sun2013,Tunali2015}
or rings from quadratic number fields \cite{Huang2015b} can have
better computation rates for some complex channels than Gaussian integers
$\mathbb{Z}[i]$.

The second line of work is to incorporate more realistic channel models such as MIMO and block fading. MIMO C\&F and integer
forcing (IF) linear receivers were studied in \cite{Zhan2009,Zhan2014IT}.
Block fading was investigated in \cite{ElBakoury2015,Wang2016}.
Reference \cite{ElBakoury2015} analyzed the computation rates and argued that the rationale of decoding an integer combination of lattice
codewords still works to some extent in block fading channels. Actual implementation
of this idea based on root-LDA lattices  was later investigated in \cite{Wang2016},
where full diversity was shown for two-way relay channels and multiple-hop
line networks. As the channel coefficients in different
fading blocks are not the same, it seems natural to employ different
integer coefficients across different blocks so as to enjoy better
quantizing performance, rather than approaches of \cite{ElBakoury2015,Wang2016}
that fix the integer coefficients for the whole duration of a codeword.
However, the resulted combination may no longer be a lattice codeword,
which draws us into a dilemma.

In \cite{Huang2016a}, it was briefly suggested that number-field
constructions as in \cite{Huang2015b,CampelloLingBelfiore2016} could
be advantageous for C\&F in a block-fading scenario. Here we provide
a detailed analysis on its decoding error performance and rates.
Specifically, with these codes, the coefficients of an equation
belong to a ring, whereas $\mathbb{Z}$ is only a special case where
its conjugates are the same. This type of lattices naturally suits
block fading channels as algebraic lattice codes can be
capacity-achieving for compound block fading channels
\cite{CampelloLingBelfiore2016}. The contribution of this work is to
demonstrate the error and rates advantages of algebraic lattices for
C\&F in block fading channels, and to present a practical algorithm
to find equations with high rates.

The rest of this paper is organized as follows. In Section II, we
review some background about C\&F and algebraic number theory. In
Sections III and IV, we present our coding scheme and the analysis of error probability and achievable rates, respectively. Section V gives a search algoirthm, and the last
section provides some simulation results.

Due to the space limit, we omit some technical proofs, especially those of the closure of an algebraic lattice under $\mathcal{O}$-linear combinations and of quantization goodness of algebraic lattices. These will be provided in a forthcoming journal paper.

Notation: Matrices and column vectors are denoted by uppercase and
lowercase boldface letters. $x(i)$ and $X(i,\thinspace j)$ refer
to scalars of $\mathbf{x}$ and $\mathbf{X}$ with indexes $i$ and
$i,\thinspace j$. The set of all $n\times n$ matrices with determinant
$\pm1$ and integer coefficients will be denoted by $\mathrm{GL}_{n}(\mathbb{Z})$.
We denote $\log^{+}(x)=\max(\log(x),0)$.

\section{Preliminaries}

\subsection{Compute and forward}

Consider a general real-valued AWGN network \cite{Nazer2011} with
$L$ source nodes and $M$ relays. We assume that each source node
$l$ is operating at the same rate and define the message rate as
$R_{\mathrm{mes}}=\frac{1}{n}\log(|W|)$, where $W$ is the message
space. A message $\mathbf{w}_{l}\in W$ is encoded, via a function
$\mathcal{E}(\cdot)$, into a point $\mathbf{x}_{l}\in\mathbb{R}^{T}$,
satisfying the power constraint $\left\Vert \mathbf{x}_{l}\right\Vert ^{2}\leq TP$,
where $T$ is the block length and $P$ denotes the signal to noise
ratio (SNR). The received signal at one relay is given by
\[
\mathbf{y}=\sum_{l=1}^{L}h_{l}\mathbf{x}_{l}+\mathbf{z},
\]
where the channel coefficients $\left\{h_{l}\right\}$ remain constant
over the whole time frame, and
$\mathbf{z}\sim\mathcal{N}(\mathbf{0},\mathbf{I}_{T})$.

In the C\&F scheme \cite{Nazer2011}, $\mathbf{x}_{l}$ is a lattice
point representative of a coset in the quotient
$\Lambda_f/\Lambda_c$, where $\Lambda_f$ and $\Lambda_c$ are called
the \textit{fine} and \textit{coarse} lattices. Instead of directly
decoding the messages, a relay searches for an integer combination of
$\mathbf{w}_{l}$, $l=1,\ldots,L$. To this purpose, the relay first
estimates a linear combination of lattice codewords
$\hat{\mathbf{x}}=[\mathcal{Q}(\alpha\mathbf{y})]\thinspace\mathrm{mod}\thinspace\Lambda_c=\sum_{l=1}^{L}a_{l}\mathbf{x}_{l}$,
where $\alpha\in\mathbb{R}$ is a minimum mean square error (MMSE)
constant, and $\mathcal{Q}(\cdot)$ is a nearest neighbor quantizer to
$\Lambda_f$. For certain coding schemes, there exists an isomorphic
mapping $g(\cdot)$ between the lattice cosets $\Lambda_f/\Lambda_c$
and the message space $W$, $g(\Lambda_f/\Lambda_c)\cong W$, which
enables the relay to forward a message
$\mathbf{u}=g(\hat{\mathbf{x}})$ in the space $W$, explicitly given
by
\begin{equation}
\mathbf{u}=\sum_{l=1}^{L}g(a_{l})\mathbf{w}_{l},\label{eq:relay message}
\end{equation}
the decoding error event of a relay given
$\mathbf{h}\in\mathbb{R}^{L}$ and $\mathbf{a}\in\mathbb{Z}{}^{L}$ as
$[\mathcal{Q}(\alpha\mathbf{y})]\thinspace\mathrm{mod}\thinspace\Lambda_c\neq\sum_{l=1}^{L}a_{l}\mathbf{x}_{l}$
for optimized $\alpha$. A computation rate is said to be achievable
at a given relay if there exists a coding scheme such that the
probability of decoding error tends to zero as $T\to\infty$. The
achievable computation rates by the C\&F protocol are given in the
following theorem.
\begin{thm}
\cite{Nazer2011} The following computation rate is achievable: 
\[
R_{\mathrm{comp}}(\mathbf{h},\thinspace\mathbf{a})=\frac{1}{2}\max_{\alpha\in\mathbb{R}}\log^{+}\left(\frac{P}{|\alpha|^{2}+P\left\Vert
\alpha\mathbf{h}-\mathbf{a}\right\Vert ^{2}}\right).
\]

\end{thm}

\subsection{Number fields and algebraic lattices}

A number field is a field extension $\mathbb{K}=\mathbb{Q}(\zeta)$
that defines a minimum field containing both $\mathbb{Q}$ and a primitive
element $\zeta$. The degree of the minimum polynomial of $\zeta$,
denoted by $n$, is called the degree of $\mathbb{K}$. Any element
in $\mathbb{K}$ can be represented by using the power basis $\{1,\zeta,\thinspace...,\thinspace\zeta^{n-1}\}$,
so that if $c\in\mathbb{K}$, then $c=c_{1}+c_{2}\zeta+\ldots+c_{n}\zeta^{n-1}$
with $c_{i}\in\mathbb{Q}$. A number is called an algebraic integer
if its minimal polynomial has integer coefficients. Let $\mathbb{S}$
be the set of algebraic integers, then the integer ring is $\mathcal{O}_{\mathbb{K}}=\mathbb{K}\cap\mathbb{S}$.
For instance, $\mathbb{\mathbb{K}=Q}(\sqrt{5})$ is a quadratic field,
its power basis is $\{1,\thinspace\sqrt{5}\}$, and an integral basis
for $\mathcal{O}_{\mathbb{K}}$ is $\{1,\thinspace\frac{1+\sqrt{5}}{2}\}$.

An ideal $\mathfrak{I}$ of $\mathcal{O}_{\mathbb{K}}$ is a nonempty
subset of $\mathcal{O}_{\mathbb{K}}$ that has the following properties.
1) $c_{1}+c_{2}\in\mathfrak{I}$ if $c_{1},\thinspace c_{2}\in\mathfrak{I}$;
2) $c_{1}c_{2}\in\mathfrak{I}$ if $c_{1}\in\mathfrak{I},\thinspace c_{2}\in\mathcal{O}_{\mathbb{K}}$.
Every ideal of $\mathcal{O}_{\mathbb{K}}$ can be decomposed into
a product of prime ideals. Let $p$ be a rational prime, we have $p\mathcal{O}_{\mathbb{K}}=\prod_{i=1}^{g}\mathfrak{p}_{i}^{e_{i}}$
in which $e_{i}$ is the ramification index of prime ideal $\mathfrak{p}_{i}$.
The inertial degree of $\mathfrak{p}_{i}$ is defined as $r_{i}=[\mathcal{O}_{\mathbb{K}}/\mathfrak{p}_{i}\thinspace:\thinspace\mathbb{Z}/p\mathbb{Z}]$,
and it satisfies $\sum_{i=1}^{g}e_{i}r_{i}=n$. Each prime ideal $\mathfrak{p}_{i}$
is said to be lying above $p$.

We follow \cite{Huang2015b,CampelloLingBelfiore2016,Kositwattanarerk2015}
to build lattices by construction A over rings. Choose $\mathfrak{p}$
lying above $p$ with inertial degree $r$, so that $\mathcal{O}_{\mathbb{K}}/\mathfrak{p}\cong\mathbb{F}_{p^{r}}$.
Let $\mathbf{G}$ be a generator matrix of a $(T,\thinspace t)$ linear
code over $\mathbb{F}_{p^{r}}$ and $t<T$. An algebraic lattice $\Lambda^{\mathcal{O}_{\mathbb{K}}}(\mathcal{C})$
is generated via the following procedures.

1) Construct a codebook $\mathcal{C}=\left\{ \mathbf{x=}\mathbf{G}\mathbf{c}\thinspace\mid\mathbf{c}\in\mathbb{F}_{p^{r}}^{t}\right\} $
with multiplication over $\mathbb{F}_{p^{r}}$.

2) Define a component-wise ring isomorphism $\mathcal{M}:\thinspace\mathbb{F}_{p^{r}}\rightarrow\mathcal{O}_{\mathbb{K}}/\mathfrak{p}$,
so that $\mathcal{C}$ is mapped to the coset leaders of ${\mathcal{O}_{\mathbb{K}}/\mathfrak{p}}^{T}$
defined by $\Lambda^{*}\triangleq\mathcal{M}(\mathcal{C})$.

3) Expand $\Lambda^{*}$ by tiling
$\Lambda^{\mathcal{O}_{\mathbb{K}}}(\mathcal{C})=\Lambda^{*}+\mathfrak{p}^{T}$.

Since $\Lambda^{\mathcal{O}_{\mathbb{K}}}(\mathcal{C})$ is an $\mathcal{O}_{\mathbb{K}}$-module
of rank $T$, a summation over $\mathcal{O}_{\mathbb{K}}$ is closed
in this group.

\section{Algebraic coding for block fading channels}

\begin{figure}[t]
\center

\includegraphics[scale=0.25]{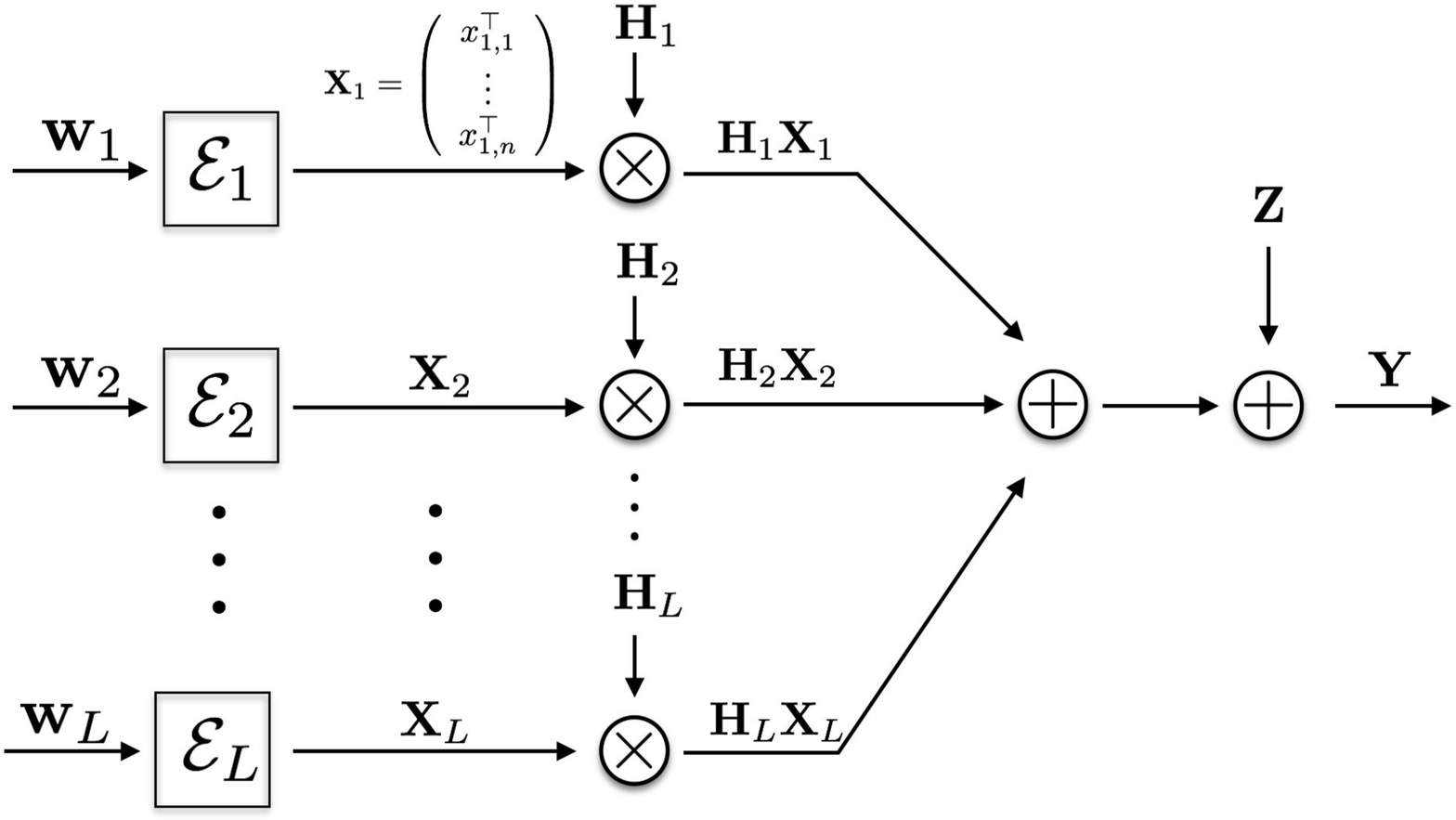}

\protect\caption{The block fading model at one relay.}
\label{fig8 block MAC}
\end{figure}

For a block fading scenario consisting of $n$ blocks and block length
$T$, the received message in a relay written in a matrix format is
\begin{equation}
\mathbf{Y}=\sum_{l=1}^{L}{\mathbf{H}}_{l}\mathbf{X}_{l}+\mathbf{Z},\label{eq: fad3}
\end{equation}
where the channel state information (CSI) ${\mathbf{H}}_{l}=\mathrm{diag}(h_{l,1},\thinspace...,h_{l,n})$
is available at the relay, $\mathbf{X}_{l}=[\mathbf{x}_{l,1},\thinspace...,\mathbf{x}_{l,n}]^{\top}\in\mathbb{R}^{n\times{T}}$
denotes a transmitted codeword, and $\mathbf{Z}=[\mathbf{z}_{1},\thinspace...,\mathbf{z}_{n}]^{\top}$
with $\mathbf{z}_{i}\sim\mathcal{N}(\mathbf{0},\thinspace\mathbf{I}_{T})$
being Gaussian noise. A diagram for this block fading channel model
is shown in Fig. \ref{fig8 block MAC}. In this figure, each $\mathbf{X}_{l}$
consists of codes over multiple frequency carriers or multiple antennas.
If our channel matrices $\left\{ {\mathbf{H}}_{l}\right\} $ are not
restricted to be diagonal, then the general model is called MIMO C\&F
\cite{Zhan2009}.

In our transmission scheme, an $\mathcal{O}_{\mathbb{K}}$-module
of rank $T$ is built first, where the degree of $\mathbb{K}$ matches
the size of the block fading channel. The coding lattice is however
not $\Lambda^{\mathcal{O}_{\mathbb{K}}}(\mathcal{C})$ as that of
\cite{Huang2015b}, but rather its canonical embedding into the Euclidean
space defined as $\Lambda^{\mathbb{Z}}(\mathcal{C})$, which is a
free $\mathbb{Z}$-module of rank $nT$. The canonical embedding is
$\sigma:\thinspace\mathbb{K}\rightarrow\mathbb{R}^{n}$, where $\sigma(x)=(\sigma_{1}(x),\thinspace...,\sigma_{n}(x))$
and all the embeddings are real. $\sigma_{1}(x),\thinspace...,\sigma_{n}(x)$
are also called the conjugates of $x$, and the algebraic norm of
$x$ is $\mathrm{Nr}(x)=\prod_{i=1}^{n}\sigma_{i}(x)$. The generator
matrix of $\Lambda^{\mathbb{Z}}(\mathcal{C})$ can be found in \cite[Prop. 1]{Kositwattanarerk2015}.

First we construct a pair of linear codes $(\mathcal{C}_{f},\thinspace\mbox{\ensuremath{\mathcal{C}}}_{c})$
to build the coding lattice $\Lambda_{f}^{\mathbb{Z}}$ and the shaping
lattice $\Lambda_{c}^{\mathbb{Z}}$. Define $\mathcal{C}_{f}=\left\{ \mathbf{G}_{f}\mathbf{w}\mid\mathbf{w}\in\mathbb{F}_{p^{r}}^{l_{f}}\right\} $
and $\mathcal{C}_{c}=\left\{ \mathbf{G}_{c}\mathbf{w}\mid\mathbf{w}\in\mathbb{F}_{p^{r}}^{l_{c}}\right\} ,$
where $\mathbf{G}_{f}\in\mathbb{F}_{p^{r}}^{T\times l_{f}}$ and $\mathbf{G}_{c}\in\mathbb{F}_{p^{r}}^{T\times l_{c}}$
is contained in the first $l_{c}$ columns of $\mathbf{G}_{f}$. Then
the fine and coarse lattices are given by $\Lambda_{f}^{\mathcal{O}_{\mathbb{K}}}=\mathcal{M}(\mathcal{C}_{f})+\mathfrak{p}^{T}$
and $\Lambda_{c}^{\mathcal{O}_{\mathbb{K}}}=\mathcal{M}(\mathcal{C}_{c})+\mathfrak{p}^{T}$.
For the time being, a candidate lattice code $\widetilde{\mathbf{x}}_{l}$
belongs to $\Lambda_{f}^{\mathcal{O}_{\mathbb{K}}}\cap\mathcal{V}(\Lambda_{c}^{\mathcal{O}_{\mathbb{K}}})$.
Since $[\mathbb{K}:\mathbb{Q}]=n$, we generate a transmitted vector
by the canonical embedding, i.e., $\mathbf{x}_{l}=\gamma\sigma(\widetilde{\mathbf{x}}_{l})\in\mathbb{R}^{nT}$,
and $\gamma$ denotes a scaling constant such that the second moment
of the shaping lattice $\gamma\mathcal{V}(\Lambda_{c}^{\mathbb{Z}})$
has a power smaller than $P$. Now we have $\mathbf{x}_{l}\in\gamma\Lambda_{f}^{\mathbb{Z}}\cap\gamma\mathcal{V}(\Lambda_{c}^{\mathbb{Z}})$.
By rearranging $\mathbf{x}_{l}$ into $\mathbf{X}_{l}$, it represents
the row composition of the conjugates of $\widetilde{\mathbf{x}}_{l}$,
i.e.,
\[
\mathbf{X}_{l}=\gamma\left[\begin{array}{c}
\sigma_{1}(\widetilde{\mathbf{x}}_{l}^{\top})\\
\sigma_{2}(\widetilde{\mathbf{x}}_{l}^{\top})\\
\vdots\\
\sigma_{n}(\widetilde{\mathbf{x}}_{l}^{\top})
\end{array}\right].
\]
Similar to \cite[Thm. 5]{Huang2015b}, there exists an isomorphism
between
$\gamma\Lambda_{f}^{\mathbb{Z}}/\gamma\Lambda_{c}^{\mathbb{Z}}$ and
the message space $W$. The equivalent lattices of
$\Lambda_{f}^{\mathbb{Z}}$ and $\Lambda_{c}^{\mathbb{Z}}$ have
volumes $p^{(T-l_{f})r}\gamma^{nT}\Delta_{\mathbb{K}}^{T/2}$ and
$p^{(T-l_{c})r}\gamma^{nT}\Delta_{\mathbb{K}}^{T/2}$
($\Delta_{\mathbb{K}}$ is the discriminant of $\mathbb{K}$), so the
message rate at every node is
$R_{\mathrm{mes}}=\frac{(l_{f}-l_{c})r}{T}\log(p)$.

\section{Error probability and rate analysis}

The following lemma is the crux of our decoding algorithm, which means
the rows of $\mathbf{X}_{l}$ are not only closed in $\gamma\Lambda_{f}^{\mathbb{Z}}$
under $\mathbb{Z}$-linear combinations, but more generally under $\mathcal{O}_{\mathbb{K}}$-linear combinations.
\begin{lem}
\label{lem: lattice closure}Let $a_{l}\in\mathcal{O}_{\mathbb{K}}$,
and ${\mathbf{A}}_{l}=\mathrm{diag}(\sigma_{1}(a_{l}),\thinspace...,\thinspace\sigma_{n}(a_{l}))$
for $1\leq l\leq L$, The physical layer codewords are closed under
the action of ring elements, i.e., $\sum_{l=1}^{L}\Big({\mathbf{A}}_{l}\mathbf{X}_{l}\Big)\in\gamma\Lambda_{f}^{\mathbb{Z}}.$
\end{lem}
Based on Lemma \ref{lem: lattice closure}, the decoder for block
fading channel (\ref{eq: fad3}) extracts an algebraic combination
of lattice codewords:
\begin{equation}
{\mathbf{B}}\mathbf{Y}=\underset{\mathrm{effective\thinspace codeword}}{\underbrace{\sum_{l=1}^{L}{\mathbf{A}}_{l}\mathbf{X}_{l}}}+\underset{\mathrm{effective\thinspace noise}}{\underbrace{{\mathbf{B}}\sum_{l=1}^{L}{\mathbf{H}}_{l}\mathbf{X}_{l}-\sum_{l=1}^{L}{\mathbf{A}}_{l}\mathbf{X}_{l}+{\mathbf{B}}\mathbf{Z}}},\label{eq: deeffect}
\end{equation}
where ${\mathbf{B}}=\mathrm{diag}(b_{1},\thinspace...,\thinspace b_{n}),$
$b_{i}\in\mathbb{R}$ is a constant diagonal matrix, to be optimized
in the sequel. The following proposition uses a union bound argument
to evaluate the decoding error probability w.r.t. model (\ref{eq: deeffect}),
whose proof can be found in the appendix.
\begin{prop}
\label{prop:prop1}Let $\mathbf{a}=[a_{1},\ldots,a_{L}]^{\top}\in\mathcal{O}_{\mathbb{K}}^{L}$
and keep the notation as above. The error probability of minimum-distance lattice decoding
associated to coefficient vector $\mathbf{a}$ is upper bounded
as
\begin{equation}
P_{e}({\mathbf{B}},\thinspace{\mathbf{a}})\leq\sum_{\mathbf{x}\in\Lambda_{f}^{\mathbb{Z}}\backslash\Lambda_{c}^{\mathbb{Z}}}\frac{1}{2}\exp\left(-\frac{n\Big(d_{n,T}(\gamma\mathbf{x})\Big)^{1/n}}{8\sum_{j=1}^{n}\nu_{\mathrm{eff},\thinspace j}^{2}}\right),\label{eq:error prob}
\end{equation}
where{\setlength{\belowdisplayskip}{0pt} \setlength{\belowdisplayshortskip}{0pt} \setlength{\abovedisplayskip}{0pt} \setlength{\abovedisplayshortskip}{0pt}

\[
\nu_{\mathrm{eff},\thinspace j}^{2}=|b_{j}|^{2}+P\left\Vert b_{j}\mathbf{h}_{j}-\sigma_{j}(\mathbf{a})\right\Vert ^{2},
\]

\[
\mathbf{h}_{j}\triangleq[H_{1}(j,\thinspace j),\thinspace\ldots,{H}_{L}(j,\thinspace j)]^{\top}\in\mathbb{R}^{L},
\]

\[
\sigma_{j}(\mathbf{a})=[{A}_{1}(j,\thinspace j),\thinspace\ldots,{A}_{L}(j,\thinspace j)]^{\top},
\]
 }
and
\[
d_{n,T}(\mathbf{x})\triangleq\prod_{j=1}^{n}\left(\sum_{i=(j-1)T+1}^{jT}x(i)^{2}\right)
\]
is the block-wise product distance of a lattice point $\mathbf{x}$.
\end{prop}
Further define the minimum block-wise product distance of a lattice
as $d_{\mathrm{min}}(\Lambda)\triangleq\min_{\mathbf{x}\in\Lambda\backslash\mathbf{0}}d_{n,T}(\mathbf{x})$.
It follows from (\ref{eq:error prob}) that the decoding error probability
is dictated by $d_{\mathrm{min}}(\gamma\Lambda_{f}^{\mathbb{Z}})$
and the power of the effective noise. The first advantage of coding
over algebraic lattices is to bring a lower bound to $d_{\mathrm{min}}(\gamma\Lambda_{f}^{\mathbb{Z}}).$
To be concise, we have $\mathrm{Nr}(x(i))\in\mathbb{Z}$ for $x(i)\in\mathcal{O}_{\mathbb{K}}$,
so that for a $\gamma\mathbf{x}\in\gamma\Lambda_{f}^{\mathbb{Z}}\neq\mathbf{0}$,
it yields
\begin{align*}
d_{n,T}(\gamma\mathbf{x}) & =\gamma^{2n}\prod_{j=1}^{n}\Big(\sum_{i=(j-1)T+1}^{jT}x(i)^{2}\Big),\\
 & \geq\gamma^{2n}\prod_{j=1}^{n}\Big(T\Big(\prod_{i=(j-1)T+1}^{jT}x(i)^{2}\Big)^{1/T}\Big),\\
 & =\gamma^{2n}T^{n}\Big(\prod_{i=1}^{T}\mathrm{Nr}(x(i))^{2}\Big)^{1/T}\geq\gamma^{2n}T^{n}.
\end{align*}

The second advantage of our scheme is that it often yields smaller
effective noise power due to finer quantization than
$\mathbb{Z}^{L}$. This will be reflected by the computation rate
analysis. According to the proof of Proposition \ref{prop:prop1}, the
nub to obtain the computation rate hinges on decoding the fine
lattice under the block-wise additive noise. Define
$\sigma_{\mathrm{GM}}^{2}=\left(\prod_{j=1}^{n}\nu_{\mathrm{eff},\thinspace
j}^{2}\right)^{1/n}$. It follows from \cite[Thm.
2]{CampelloLingBelfiore2016} that the decoding error probability
vanishes if
$\mathrm{Vol}(\gamma\Lambda_{f}^{\mathbb{Z}})^{2/(nT)}/\sigma_{\mathrm{GM}}^{2}>
2\pi e$. From quantization goodness \cite{Huang2015b}, we have that
$P/\mathrm{Vol}(\gamma\Lambda_{c}^{\mathbb{Z}})^{2/(nT)} < 1/(2\pi
e)$. Therefore, any computation rate up to
\begin{eqnarray*}
\frac{1}{T}\log\left(\frac{\mathrm{Vol}(\gamma\Lambda_{c}^{\mathbb{Z}})}{\mathrm{Vol}(\gamma\Lambda_{f}^{\mathbb{Z}})}\right) & \leq & \frac{n}{2}\log\left(\frac{P}{\sigma_{\mathrm{GM}}^{2}}\right)
\end{eqnarray*}
is achievable. In order to relate the achievable rate to the
successive minima of a lattice, we define
$\sigma_{\mathrm{AM}}^{2}=\left(\sum_{j=1}^{n}\nu_{\mathrm{eff},\thinspace
j}^{2}\right)/n$ and use the fact that
$\sigma_{\mathrm{AM}}^{2}\geq\sigma_{\mathrm{GM}}^{2}$ to attain the
following result.
\begin{prop}
\label{thm:rate} With properly chosen lattice codebooks, given channels
$\left\{ {\mathbf{H}}_{l}\right\} $ and the desired quantization
coefficient $\mathbf{a}$ in a relay, the computation rate of the
arithmetic mean (AM) decoder is given by
\begin{multline}
R_{\mathrm{AM}}(\left\{ {\mathbf{H}}_{l}\right\}, \mathbf{a})=\\
\max_{{\mathbf{B}}}\frac{n}{2}\log^{+}\left(\frac{nP}{\left\Vert {\mathbf{B}}\right\Vert ^{2}+P\sum_{l=1}^{L}\left\Vert {\mathbf{B}}{\mathbf{H}}_{l}-{\mathbf{A}}_{l}\right\Vert ^{2}}\right).\label{eq:am}
\end{multline}
\end{prop}

Denote the denominator inside (\ref{eq:am}) as
$n\sigma_{\mathrm{AM}}^{2}=\left\Vert {\mathbf{B}}\right\Vert
^{2}+P\sum_{l=1}^{L}\left\Vert
{\mathbf{B}}{\mathbf{H}}_{l}-{\mathbf{A}}_{l}\right\Vert ^{2}$. By
assuming $\mathbf{a}$ to be fixed, the MMSE principle for optimizing
$n\sigma_{\mathrm{AM}}^{2}$ is to pick the diagonal elements of
$\mathbf{B}$ in the following way:
\begin{equation}
b_{j}=\frac{P\sigma_{j}(\mathbf{a})^{\top}\mathbf{h}_{j}}{P\left\Vert \mathbf{h}_{j}\right\Vert ^{2}+1}.\label{eq:b coeff}
\end{equation}
Plugging (\ref{eq:b coeff}) back yields
\begin{multline*}
n\sigma_{\mathrm{AM}}^{2}\\
=\sum_{j=1}^{n}\left(P\left\Vert \frac{P\mathbf{h}_{j}\sigma_{j}(\mathbf{a})^{\top}\mathbf{h}_{j}}{P\left\Vert \mathbf{h}_{j}\right\Vert ^{2}+1}-\sigma_{j}(\mathbf{a})\right\Vert ^{2}+\left(\frac{P\sigma_{j}(\mathbf{a})^{\top}\mathbf{h}_{j}}{P\left\Vert \mathbf{h}_{j}\right\Vert ^{2}+1}\right)^{2}\right).
\end{multline*}
Further define a Gram matrix
\[
\mathbf{M}_{j}=\mathbf{I}-\frac{P}{P\left\Vert \mathbf{h}_{j}\right\Vert ^{2}+1}\mathbf{h}_{j}\mathbf{h}_{j}^{\top},
\]
then the computation rate of our AM decoder becomes

\begin{equation}
R_{\mathrm{AM}}\left(\left\{ \mathbf{M}_{j}\right\},
\mathbf{a}\right)=\frac{n}{2}\log^{+}\left(\frac{n}{\sum_{j=1}^{n}\sigma_{j}(\mathbf{a})^{\top}\mathbf{M}_{j}\sigma_{j}(\mathbf{a})}\right).\label{eq:rate
am2}
\end{equation}
Its achievable rate is therefore maximized by optimizing $\mathbf{a}\in\mathcal{O}_{\mathbb{K}}^{L}$.
Since $\mathbb{Z}^{L}\subseteq\mathcal{O}_{\mathbb{K}}^{L}$, the
achievable rate in (\ref{eq:rate am2}) is no smaller than that of
$\mathbb{Z}$-lattices.

\section{Search algorithm}

The optimization target in (\ref{eq:rate am2}) is to find $\mathbf{a}\in\mathcal{O}_{\mathbb{K}}^{L}$
to reach the minimum of $f(\mathbf{a})\triangleq\sum_{j=1}^{n}\sigma_{j}(\mathbf{a})^{\top}\mathbf{M}_{j}\sigma_{j}(\mathbf{a})$.
Our approach is to take advantage of the generator matrix of $\mathcal{O}_{\mathbb{K}}$,
so that $f(\mathbf{a})$ represents the square distance of a lattice
vector, and (\ref{eq:rate am2}) is turned into a shortest vector
problem (SVP). Let $\left\{ \phi_{1},\thinspace...,\thinspace\phi_{n}\right\} $
be a $\mathbb{Z}$-basis of $\mathcal{O}_{\mathbb{K}}$, then its
generator matrix is given by
\[
\Phi=\left[\begin{array}{ccc}
\sigma_{1}(\phi_{1}) & \cdots & \sigma_{1}(\phi_{n})\\
\sigma_{2}(\phi_{1}) & \cdots & \sigma_{2}(\phi_{n})\\
\vdots & \vdots & \vdots\\
\sigma_{n}(\phi_{1}) & \cdots & \sigma_{n}(\phi_{n})
\end{array}\right].
\]
With Cholesky decomposition $\mathbf{M}_{j}=\bar{\mathbf{M}}_{j}^{\top}\bar{\mathbf{M}}_{j}$,
we have $f(\mathbf{a})=\sum_{j=1}^{n}\left\Vert \bar{\mathbf{M}}_{j}\sigma_{j}(\mathbf{a})\right\Vert ^{2}$.
The lattice associated with $f(\mathbf{a})$ is indeed a $\mathbb{Z}$-submodule
of $\mathbb{R}^{nL}$, with a generator matrix $\bar{\Phi}=\mathbf{M}_{\mathrm{mix}}\Phi_{\mathrm{mix}}$,
\[
\mathbf{M}_{\mathrm{mix}}=\left[\begin{array}{ccc}
\bar{\mathbf{M}}_{1} & \cdots & \mathbf{0}\\
\mathbf{0} & \cdots & \mathbf{0}\\
\vdots & \vdots & \vdots\\
\mathbf{0} & \cdots & \bar{\mathbf{M}}_{n}
\end{array}\right],
\]
and $\Phi_{\mathrm{mix}}=\mathbf{U}(\mathbf{I}_{L}\otimes\Phi)$ where
$\mathbf{U}\in\mathrm{GL}_{nL}(\mathbb{Z})$ is a row-shuffling operation.
For instance, when $n=2,\thinspace L=2$, we can visualize $\Phi_{\mathrm{mix}}$
as
\[
\Phi_{\mathrm{mix}}=\left[\begin{array}{cccc}
\sigma_{1}(\phi_{1}) & \sigma_{1}(\phi_{2}) & 0 & 0\\
0 & 0 & \sigma_{1}(\phi_{1}) & \sigma_{1}(\phi_{2})\\
\sigma_{2}(\phi_{1}) & \sigma_{2}(\phi_{2}) & 0 & 0\\
0 & 0 & \sigma_{2}(\phi_{1}) & \sigma_{2}(\phi_{2})
\end{array}\right].
\]
Finally, it yields $f(\mathbf{a})=f(\tilde{\mathbf{a}})=\left\Vert \bar{\Phi}\tilde{\mathbf{a}}\right\Vert ^{2}$,
with $\tilde{\mathbf{a}}\in\mathbb{Z}^{nL}$. Many algorithms are
now available to solve SVP over the $\mathbb{Z}$ lattice $\mathcal{L}(\bar{\Phi})$,
e.g., the classic sphere decoding algorithm \cite{Schnorr1994} can
help to obtain this solution with reasonable complexity.

The explicit structure of lattice basis $\bar{\Phi}$ facilitates
estimating the bounds of rates via different number fields. Denote
the first successive minimum of $\mathcal{L}(\bar{\Phi})$ by
$\lambda_{1}$, we have
$\lambda_{1}<\sqrt{nL}|\det\left(\bar{\Phi}\right)|^{1/\left(nL\right)}$
according to Minkowski's first theorem \cite[P. 12]{Micciancio2002}.
We claim that a smaller discriminant $\Delta_{\mathbb{K}}$ can
contribute to a sharper bound for it, so that $\mathbb{Q}(\sqrt{5})$
should be the best real quadratic number field to use. Specifically,
$|\det(\bar{\Phi})|=|\det(\mathbf{M}_{\mathrm{mix}})||\det(\Phi_{\mathrm{mix}})|$,
and since
$|\det(\Phi_{\mathrm{mix}})|=|\det(\Phi)|^{L}=\left(\Delta_{\mathbb{K}}\right)^{L/2}$due
to the unimodularity of $\mathbf{U}$, it yields
$|\det(\bar{\Phi})|=|\det(\mathbf{M}_{\mathrm{mix}})|\left(\Delta_{\mathbb{K}}\right)^{L/2}$.
The relation to channel capacity can also be obtained by using
Sylvester\textquoteright s Theorem to expand each
$|\det(\bar{\mathbf{M}}_{i})|$ as Ref. \cite{Nazer2016} did to the
static Gaussian MAC.

\section{Numerical results}

In this section, we will numerically verify the validness of the AM
computation rate (\ref{eq:rate am2}). In the example, we let $n=2$,
$L=2$, $\mathbf{h}_{1}$ and \textbf{$\mathbf{h}_{2}$ }chosen from
$\mathcal{N}(0,\thinspace1)$ entries, and compare the average
achievable rates (ergodic rates) of $2000$ Monte Carlo runs.

\begin{figure}[t]
\center

\includegraphics[clip,width=3.2in]{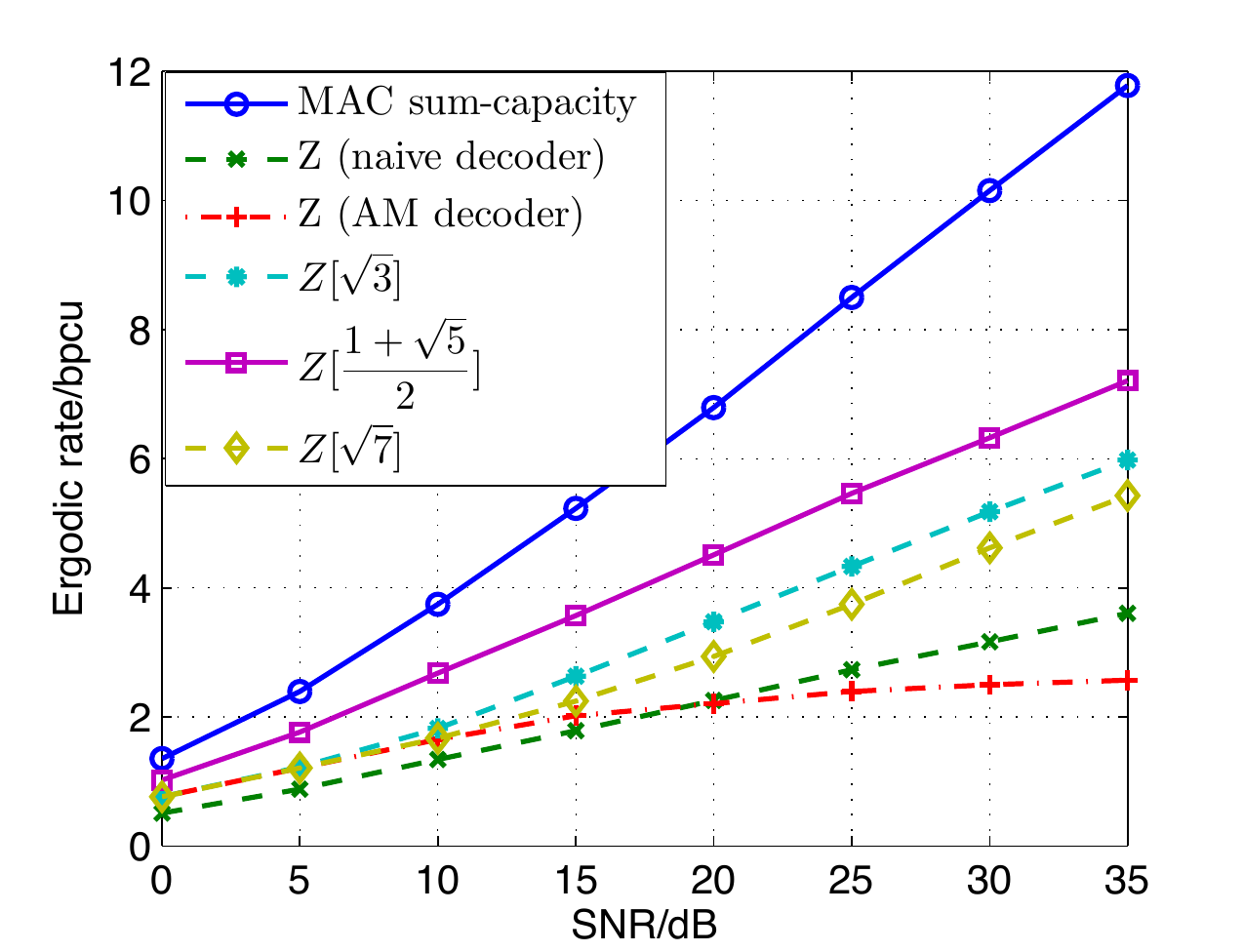}

\protect\caption{Comparison of achievable rates with different rings.}
\label{fig8 block MAC-1}
\end{figure}

In Fig. \ref{fig8 block MAC-1}, we plot the rates of AM decoders
with quantization coefficients in $\mathbb{Z}$, $\mathbb{Z}[\sqrt{3}]$,
$\mathbb{Z}[\frac{1+\sqrt{5}}{2}]$, and $\mathbb{Z}[\sqrt{7}]$,
respectively. The MAC sum-capacity is provided as the upper bound
of decoding two equations. The rate of an oblivious transmitter \cite{Zhan2009}
that neglects the advantage of multiple antennas is also included
in the figure, denoted as $\mathbb{Z}$ (naive decoder).

We can observe from Fig. \ref{fig8 block MAC-1} that the degree of
freedom (DOF) of the MAC sum-capacity is $2$, the DOF's of non-trivial
rings are 1 (and they are optimal because decoding two equations suffices
to reach the DOF $2$), and that of the naive decoder is only $1/2$.
The performance of $\mathbb{Z}[\frac{1+\sqrt{5}}{2}]$ is better than
those of other rings. The AM decoder with the $\mathbb{Z}$ restriction
seems quite sub-optimal, as it becomes inferior to the naive decoder
in high SNR.

\appendices{}

\section{\label{sec: short quan}Proof of Proposition \ref{prop:prop1}}
\begin{IEEEproof}
\noindent We first follow \cite{Nazer2011} to find the effective
noise. With chosen ${\mathbf{B}}$ and ${\mathbf{A}}_{l}$, it first
computes $\mathbf{S}={\mathbf{B}}\mathbf{Y}+\sum_{l=1}^{L}{\mathbf{A}}_{l}\mathbf{D}_{l}$,
where $\mathbf{D}_{l}$ is the dither from a source node which is
uniformly distributed on the Voronoi region $\gamma\mathcal{V}_{\Lambda_{c}^{\mathbb{Z}}}$.
To get an estimate of the lattice equation \textbf{$\mathbf{V}=\sum_{l=1}^{L}{\mathbf{A}}_{l}\mathbf{X}_{l}\thinspace\mod\thinspace\gamma\Lambda_{c}^{\mathbb{Z}}$},
$\mathbf{S}$ is first quantized w.r.t. the fine lattice $\gamma\Lambda_{f}^{\mathbb{Z}}$
denoted by $\mathcal{Q}(\cdot)$ and then modulo the coarse lattice
$\gamma\Lambda_{c}^{\mathbb{Z}}$. Since
\[
[\mathcal{Q}(\mathbf{S})]\thinspace\mod\thinspace\gamma\Lambda_{c}^{\mathbb{Z}}=[\mathcal{Q}([\mathbf{S}]\thinspace\mod\thinspace\gamma\Lambda_{c}^{\mathbb{Z}})]\thinspace\mod\thinspace\gamma\Lambda_{c}^{\mathbb{Z}},
\]
if the effective noise of $[\mathbf{S}]\thinspace\mod\thinspace\gamma\Lambda_{c}^{\mathbb{Z}}$
falls within the Voronoi region of the fine lattice,
then the noise effect can be canceled. Now we show that $[\mathbf{S}]\thinspace\mod\thinspace\gamma\mathcal{V}_{\Lambda_{c}^{\mathbb{Z}}}$
is equivalent to $\mathbf{V}$ pluses a block-wise noise. Denote $\Theta_{l}={\mathbf{B}}{\mathbf{H}}_{l}-{\mathbf{A}}_{l}$
and $\bar{\mathbf{X}}_{l}=[\mathbf{X}_{l}+\mathbf{D}_{l}]\thinspace\mod\thinspace\gamma\Lambda_{c}^{\mathbb{Z}}$,
then {\setlength{\abovedisplayskip}{0pt} \setlength{\belowdisplayskip}{.5pt}

\begin{align*}
[\mathbf{S}]\mod\thinspace\gamma\Lambda_{c}^{\mathbb{Z}} & =[\mathbf{V}+\underset{\mathbf{Z}_{\mathrm{eff}}}{\underbrace{\sum_{l=1}^{L}({\Theta}_{l}\bar{\mathbf{X}}_{l})+{\mathbf{B}}\mathbf{Z}}}]\thinspace\mod\thinspace\gamma\Lambda_{c}^{\mathbb{Z}}.
\end{align*}
}As each block of $\bar{\mathbf{X}}_{l}$ is uniformly distributed,
the probability density function (PDF) of the $j$th row of
$\mathbf{Z}_{\mathrm{eff}}$ can be shown to be upper bounded by a
Gaussian
$\mathcal{N}(\mathbf{0},\thinspace\nu_{\mathrm{eff},\thinspace
j}^{2}\mathbf{I}_{T})$, where {\setlength{\abovedisplayskip}{0pt}
\setlength{\belowdisplayskip}{.5pt}

\begin{equation}
\nu_{\mathrm{eff},\thinspace j}^{2}=|b_{j}|^{2}+P\left\Vert b_{j}\mathbf{h}_{j}-\sigma_{j}(\mathbf{a})\right\Vert ^{2},\label{eq:app1}
\end{equation}
}

It turns out to be a non-AWGN lattice decoding problem, whose decoding
error probability is $P_{e}({\mathbf{B}},\thinspace\mathbf{a})=\sum_{\mathbf{V}'\in\left\{ \mathbf{V}+\gamma\Lambda_{f}^{\mathbb{Z}}\right\} \backslash\left\{ \mathbf{V}+\gamma\Lambda_{c}^{\mathbb{Z}}\right\} }\mathrm{Pr}(\mathbf{V}\rightarrow\mathbf{V}')$
which equals {\setlength{\abovedisplayskip}{.5pt} \setlength{\belowdisplayskip}{.5pt}
{\small{}
\begin{multline}
\sum_{\mathbf{V}-\mathbf{V}'\in\gamma\Lambda_{f}^{\mathbb{Z}}\backslash\gamma\Lambda_{c}^{\mathbb{Z}}}\mathrm{Pr}\Big(\left\Vert \mathbf{V}+\mathbf{Z}_{\mathrm{eff}}-\mathbf{V}'\right\Vert ^{2}\!\leq\left\Vert \mathbf{Z}_{\mathrm{eff}}\right\Vert ^{2}\Big),\\
=\sum_{\mathbf{V}-\mathbf{V}'\in\gamma\Lambda_{f}^{\mathbb{Z}}\backslash\gamma\Lambda_{c}^{\mathbb{Z}}}\mathrm{Pr}\Big(\sum_{j=1}^{n}(\left\Vert \mathbf{v}_{j}-\mathbf{v}_{j}'\right\Vert ^{2}+2(\mathbf{v}_{j}-\mathbf{v}_{j}')^{\top}\mathbf{z}_{\mathrm{eff},\thinspace j})\leq0\Big),\label{eq:sum index}
\end{multline}
}}in which $\mathbf{v}_{j}^{\top}$, $\mathbf{v}_{j}'^{\top}$ and
$\mathbf{z}_{\mathrm{eff},\thinspace j}^{\top}$ are the $j$th rows
of $\mathbf{V},$ $\mathbf{V}'$ and $\mathbf{Z}_{\mathrm{eff}}$,
respectively. Further define $\varUpsilon\triangleq\sum_{j=1}^{n}2(\mathbf{v}_{j}'-\mathbf{v}_{j})^{\top}\mathbf{z}_{\mathrm{eff},\thinspace j}$.
Similar to the analysis of (\ref{eq:app1}), the PDF of $\varUpsilon$
is upper bounded by a zero mean Gaussian with variance $\sum_{j=1}^{n}4\nu_{\mathrm{eff},\thinspace j}^{2}\left\Vert \mathbf{v}_{j}-\mathbf{v}_{j}'\right\Vert ^{2}$.
It then follows from the property of a Q function $Q_{g}(x)\triangleq\frac{1}{\sqrt{2\pi}}\int_{x}^{\infty}\exp\Big(-\frac{u^{2}}{2}\Big)\mathrm{d}u$
that the summation term of (\ref{eq:sum index}) can be written as
\begin{align}
\mathrm{Pr}(\mathbf{V}\rightarrow\mathbf{V}') & \leq Q_{g}\Bigg(\frac{\sum_{j=1}^{n}\left\Vert \mathbf{v}_{j}-\mathbf{v}_{j}'\right\Vert ^{2}}{2\sqrt{\sum_{j=1}^{n}\nu_{\mathrm{eff},\thinspace j}^{2}\left\Vert \mathbf{v}_{j}-\mathbf{v}_{j}'\right\Vert ^{2}}}\Bigg),\nonumber \\
 & \stackrel{(a)}{\leq}\frac{1}{2}\exp\Bigg(-\frac{\Big(\sum_{j=1}^{n}\left\Vert \mathbf{v}_{j}-\mathbf{v}_{j}'\right\Vert ^{2}\Big)^{2}}{8\sum_{j=1}^{n}\nu_{\mathrm{eff},\thinspace j}^{2}\left\Vert \mathbf{v}_{j}-\mathbf{v}_{j}'\right\Vert ^{2}}\Bigg),\nonumber \\
 & \stackrel{(b)}{\leq}\frac{1}{2}\exp\Bigg(-\frac{n\Big(\prod_{j=1}^{n}\left\Vert \mathbf{v}_{j}-\mathbf{v}_{j}'\right\Vert ^{2}\Big)^{1/n}}{8\sum_{j=1}^{n}\nu_{\mathrm{eff},\thinspace j}^{2}}\Bigg),\label{eq:app3}
\end{align}
where (a) has used the bound $Q_{g}(x)\leq1/2\exp(-x^{2}/2)$, (b)
comes after using $\nu_{\mathrm{eff},\thinspace j}^{2}\leq\sum_{j=1}^{n}\nu_{\mathrm{eff},\thinspace j}^{2}$
and the AM-GM inequality. The relaxation in (b) serves the purpose
of bounding the error probability via the block-wise product distance
of our algebraic lattice. Plugging (\ref{eq:app1}) into (\ref{eq:app3})
proves the proposition.
\end{IEEEproof}

\section*{acknowledgment}

The authors acknowledge Dr. Yu-Chih (Jerry) Huang for fruitful discussions.

\bibliographystyle{IEEEtranMine}
\addcontentsline{toc}{section}{\refname}\bibliography{library}

\end{document}